**Main Manuscript for**

Data-Driven Modeling Reveals the Impact of Stay-at-Home Orders on Human Mobility during the COVID-19 Pandemic in the U.S.

Chenfeng Xiong, Songhua Hu, Mofeng Yang, Hannah N. Younes, Weiyu Luo, Sepehr Ghader, Lei Zhang*

Maryland Transportation Institute (MTI)
University of Maryland, College Park, MD 20742
* Corresponding author: Lei Zhang
**Email:**  cxiong@umd.edu, hsonghua@umd.edu, mofeng@umd.edu, hyounes@terpmail.umd.edu, wyl@umd.edu, sghader@umd.edu, lei@umd.edu

**Classification**
Social Sciences

**Abstract**

One approach to delay the spread of the novel coronavirus (COVID-19) is to reduce human travel by imposing travel restriction policies. It is yet unclear how effective those policies are on suppressing the mobility trend due to the lack of the ground truth and large-scale dataset describing human mobility during the pandemic. This study uses real-world location-based service data collected from anonymized mobile devices to uncover mobility changes during COVID-19 and under the "Stay-at-home" state orders in the U.S. The study measures human mobility with two important metrics: daily average number of trips per person and daily average person-miles traveled. The data-driven analysis and modeling attributes less than 5% of the reduction in number of trips and person-miles traveled to the effect of policy. The models developed in the study exhibit high prediction accuracy and can be applied to inform epidemics modeling with empirically verified mobility trends and to support time-sensitive decision-making processes.

**Significance Statement**

This paper measures human mobility changes in the U.S. during the novel coronavirus (COVID-19) pandemic and quantifies the impact of "stay-at-home" state orders on human mobility. Our human mobility analyses draw evidences from a large-scale and anonymized nationwide location-based service data and develop a dynamic econometric model with high prediction accuracy. It deepens the understanding of human mobility during COVID-19 the control of spread of COVID-19 and can be used to support immediate decision-making that needs empirical evidences on human movement to assess transmission of disease and control measures.

**Introduction**

The coronavirus disease 2019 (COVID-19) pandemic is undoubtedly one of the worst global health crises seen in decades. The first confirmed case of COVID-19 in the U.S. emerged in Washington State on January 21st, 2020. Three months later, over three quarters of a million cases had been confirmed throughout the nation. Governments around the world are taking rapid action to mitigate the spread of the disease. The U.S. government proclaimed a national state of emergency on March 13th, 2020. Following an exponential growth in the number of confirmed cases, the Federal Emergency Management Agency (FEMA) announced its first major disaster in the state of New York on March 20th, 2020, followed by California and Washington on March 22nd (FEMA, 2020). As of April 11th, 2020, FEMA had announced the COVID-19 pandemic a disaster in every state, with Wyoming being the last one. On March 19th, 2020, California became the first state to institute a "Stay-at-home" or "Shelter in place" order (State of California, 2020). By mid-April 2020, all but 8 states had followed suit. Three of the 8 states had partial stay-at-home orders, implemented by city mayors or county executives (Mervosh et al., 2020).

One approach to delay the spread is to reduce human travel by imposing travel restriction policies. International and domestic travel restrictions are shown to possibly decrease the rate of case exportations (Wells et al., 2020; Chinazzi et al., 2020). Government appropriations are a crucial component in reducing case fatalities in the U.S. (Moghadas et al., 2020). Research focused on analyzing the impact of such policies on the spread of COVID-19 generally use epidemic models, such as GLEAM, while people's compliance to such travel restriction policies are either not assumed or assumed as a single factor (Chinazzi et al., 2020). It is still unclear how effective those policies are on suppressing the mobility trend to flatten the curve largely due to the lack of the ground truth and large-scale dataset describing human mobility patterns during the pandemic.

The authors use real-world mobility big data collected from anonymized mobile devices to uncover mobility changes under the "Stay-at-home" state orders in the U.S.. To capture the dynamic behavior response to the "Stay-at-home" state orders, a data panel of integrated and processed location data representing movements for the entire U.S. was developed, incorporating daily movements, from January 1st, 2020 to April 11th, 2020, of over 10 million anonymous and opted-in individuals daily. Then, previously developed and validated spatial-temporal algorithms (Zhang and Ghader, 2020) were used to identify all trips from the data panel. A multi-level weighting procedure expanded the sample to the entire U.S. population, using device-level and trip-level weights to ensure data representativeness in the nation, states, and counties.

The data panel and the computational algorithms have been validated based on a variety of independent datasets such as the National Household Travel Survey (NHTS) and the American Community Survey (ACS), and peer reviewed by an external expert panel in a U.S. Department of Transportation Federal Highway Administration's Exploratory Advanced Research Program project (Zhang and Ghader, 2020). Two human mobility metrics were developed at the national and state-level and integrated with COVID-19 case data collected from the Johns Hopkins University COVID-19 dashboard (Dong et al., 2020), population data, and other data sources for the analyses (ACS, 2017; FHWA, 2017).

- Daily average number of trips per person: The total number of identified trips in each day that are above three hundred meters length divided by the total population;
- Daily average person-miles traveled: The total person-miles traveled in each day across all travel modes (plane, car, bus, rail, bike, walk, etc.) divided by the total population.

**Results**

Using the human mobility metrics in January 2020 as the comparison benchmark (New Year's Day, Jan 1st and Martin Luther King Jr. Day, Jan 20th are excluded), we quantified the national mobility trend during the COVID-19, as visualized in Figure 1. People did not reduce their travels until the second week of March, during which the World Health Organization (WHO) defined COVID-19 as a global pandemic and the proclamation on declaring a national emergency was made by the White House (WHO, 2020; White House, 2020). Immediately following, the sharpest deterioration of mobility was observed in the third week of March. Mobility in terms of daily average person-miles traveled has already dropped over 20% compared to January average when the first state in the U.S., California, announced a "Stay-at-home" order. In addition, the decrease of daily average number of trips per person has not dropped as significantly, indicating that more short-distance trips were made. Those could be out-door activities, grocery, pharmacy, and fast food, as confirmed in other parallel analyses (Google, 2020; PlaceIQ, 2020).

The big-data analytics suggest that, at the national level, people were already practicing social distancing by the time "Stay-at-home" orders were issued by state governors. How effective are those orders in further limiting people's travel? By April 10th, only 8 states had not issued "Stay-at-home" orders. We reorganized the mobility data and anchored the data to the dates when different states announced their "Stay-at-home" orders, and then quantified the human mobility changes, compared to the January benchmark, three weeks, two weeks, and one week before and after the orders were issued (illustrated in Figure 2). While the data confirmed that, nationwide, mobility had dropped significantly one week or even two weeks before the orders were issued, an additional 6.1% decrease in daily average number of trips per person and 10.8% decrease in daily average person-miles traveled (PMT) were observed in the week after the order took effect across different states. The states that saw the most significant mobility drop after the orders are NJ (-17% trip, -23% PMT), NY (-17% trip, -22% PMT), IL (-16% trip, - 23% PMT), CA (-16% trip, -22% PMT), MI (-14% trip, -23% PMT). The least significant mobility drops (+-5%) can be observed in PA, KY, MO, TN, and D.C.

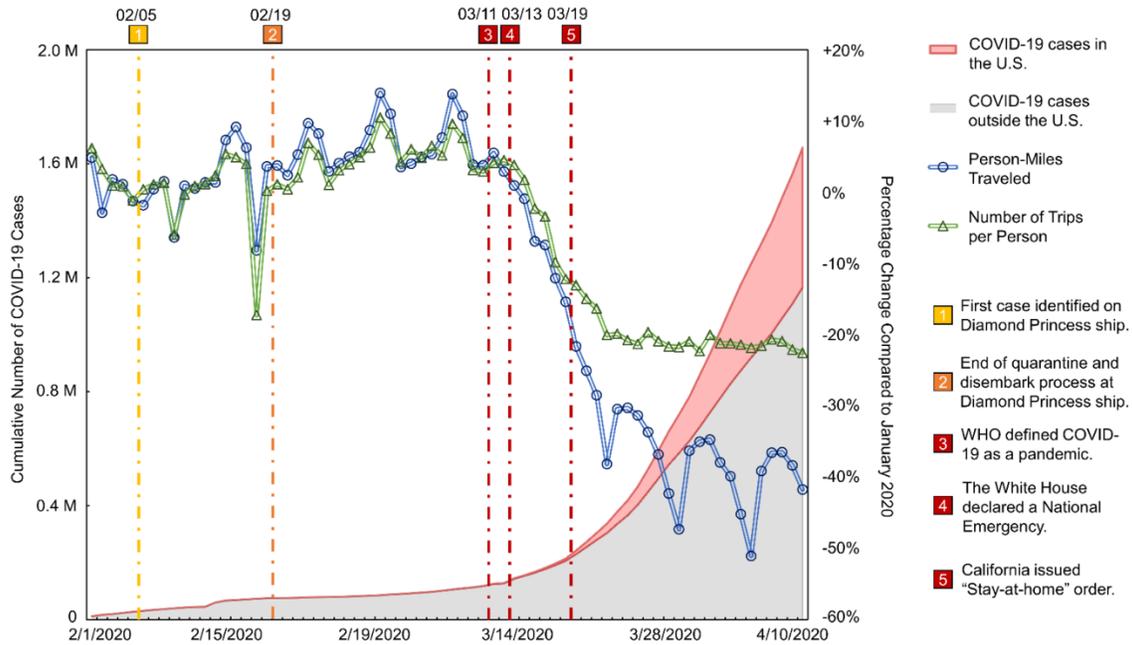

**Figure 1.** National mobility trend during the COVID-19 crisis, quantified by daily average number of trips per person and daily average person-miles traveled

"Stay-at-home" orders vary widely in when and how they are implemented and enforced. Essential activities, such as receiving healthcare services, going grocery shopping, exercising and essential work are authorized during stay-at-home orders. Differences in the orders emerge in religious gatherings, where some states allow for gatherings of more than 10 people for special events while most do not. Interstate travel is sometimes allowed, although some states, such as Alaska, require special permission to leave the state and others, like Nevada, require a 14-day quarantine period for anyone re-entering the state (State of Alaska, 2020; Fox 11, 2020). Education institutions have been ordered closed in nearly half the U.S. states for the remainder of the year. Only seven states have recommended that schools be closed as opposed to ordering closure altogether (Education Week, 2020). Seeing family members outside of their immediate household is also not permitted in several states, unless for the purpose of "caring for the family member," i.e. providing groceries for or medical aid to elderly or immunocompromised family members. Additionally, not all "Stay-at-home" orders are implemented at the state level. Many were deployed in counties or cities before the entire state adopted the order. As of mid-April, three states: Oklahoma, Utah and Wyoming had partial stay-at-home orders.

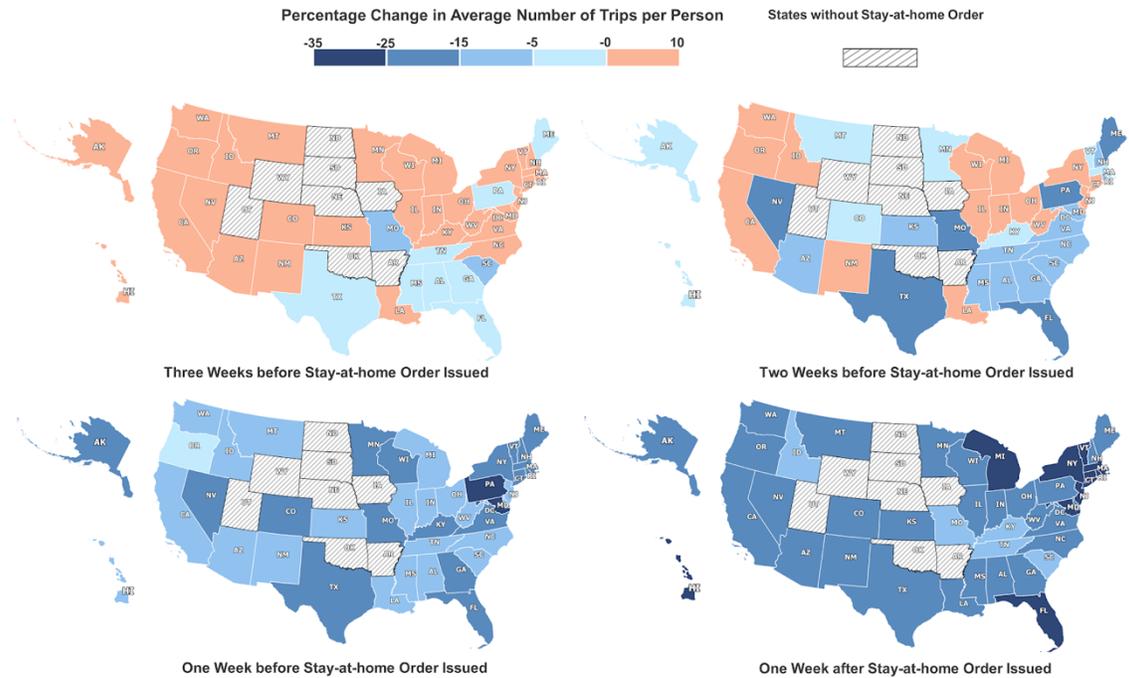

(a)

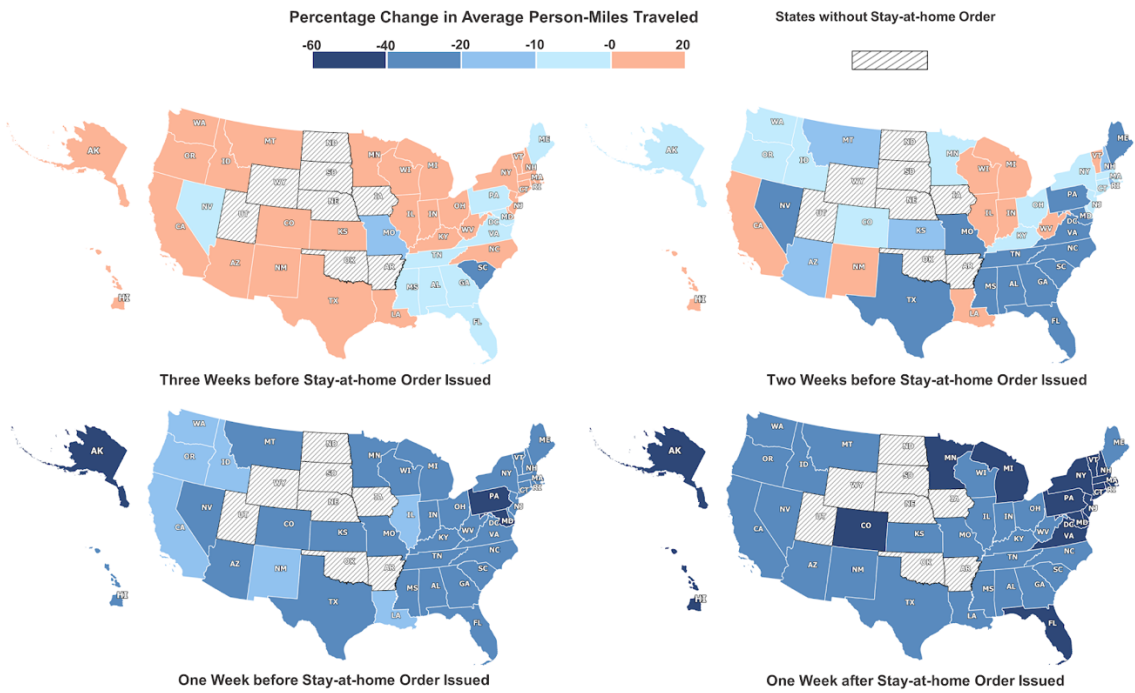

(b)

**Figure 2.** Change in (a) Average Number of Trips per Person and (b) Average Person-Miles Traveled One, Two and Three Weeks Before and One Week After a "Stay-at home" Order Took Effect Relative to January 2020

Enforcement is a crucial component of "Stay-at-home" orders. Fifteen states (including D.C.) have stated that they would issue fines and possible jail time for persons violating the order; fourteen have stated that they would enforce the order by first issuing a warning, then possibly issuing a fine for a repeated offense; and fourteen more either have not stated how the order would be enforced or have explicitly stated that it would only be enforced through education, without penalties (Mazziotta, 2020). This variation in how the order is enforced conveys a possible discrepancy in how people respond to the order. Risking a fine of $1000 or above and possible jail time in New Jersey, Hawaii, the District of Columbia, and several other states may deter residents from going out unnecessarily more so than states who solely focus on educating people who violate the order.

To quantify how people in different states responded to "Stay-at-home" orders during the COVID-19 pandemic, we studied the longitudinal changes in state-level mobility using a generalized additive model (GAM) (Wood, 2017; Hastie, 1993; Hastie & Tibshirani, 1990) of daily average number of trips per person and daily average person-miles traveled. A notable advantage of GAM lies in its flexibility that combines the linear and fixed effects with nonlinear effects such as temporal patterns, variable interaction, and individual-specific random effects. These are essential in longitudinal analysis since observations are intra-subject correlated, time-varying, and with unobserved heterogeneity. The control variables include the COVID-19 related features (daily number of newly confirmed coronavirus cases in the state and that number in the adjacent states), policy-related features (whether the order has been issued, and the levels of enforcements), and other factors contributing to the variation (i.e., random effect of time and states, weekday-weekend variation, and state governor approval rates).

Our model predicts the daily average number of trips and daily average person-miles traveled across all states with relatively high accuracy (Adj. $R^2$ = 0.882 and 0.919, respectively). Most of the parameters of control variables are found statistically significant at least at the $p<0.1$ level. As expected, the daily number of newly confirmed coronavirus cases in the state and that number for the rest of the nation are very effective in persuading people to travel less. We also find that the effect of "Stay-at-home" orders negatively contributes to human mobility. More interestingly, the magnitude of the effect intensifies as state enforcement becomes stricter.

The models are then applied to estimate the trip reduction and miles-traveled reduction due to the issuance of "Stay-at-home" orders by comparing the predictions with and without "Stay-at-home" orders. As depicted in Figure 3 and Figure 4, only a moderate amount of reductions in human mobility can be attributed to "Stay-at-home" orders. Nationally speaking, the orders lead to a daily reduction of 0.136 trips per person (or 4.9% of average daily number of trips per person) and a daily reduction of 1.185 person-miles traveled (or 4.8% of average daily person-miles traveled). State-level effect has also been quantified. Stay-at-home orders account for a greater effect in states with more strict enforcement, such as HI (the order leads to 0.21 person trips reduced and 3.55 miles reduced) and MI. Compared to states that issued Stay-at-home orders without penalty or without specifying particular enforcement, those enforced with warning and possible fines for repeated offense see an additional 0.4% in trip reduction. Enforcement with fines and possible jail time will further reduce 1.5% daily trips per person. The policy effect is greatly related to the timing of the orders. States that had an already deteriorated mobility before the order release or simply issued the order late saw a limited effect, such as NY and NJ.

Understanding and accurately predicting human mobility during a pandemic is critical for the control of spread of COVID-19 and any other highly contagious disease. Here, anonymized and privacy-enhanced location-based service data in the nation provides ground-truth empirical evidence on how people in the U.S. moved during the COVID-19 outbreak and successfully supported human mobility analytics and modeling. The models were deemed statistically significant and accurate, and estimated the effect of state-issued "Stay-at-home" orders and other influential factors on human mobility changes. Dynamics in human mobility, such as distance traveled and number of trips made in a day, were thus quantified. These model outputs can be immediately integrated in epidemics models that need empirical data support on human movement to assess transmission of disease and control measures. Our immediate next step is to determine hotspot travel destinations in decision support of mitigating spread and reducing local transmissions.

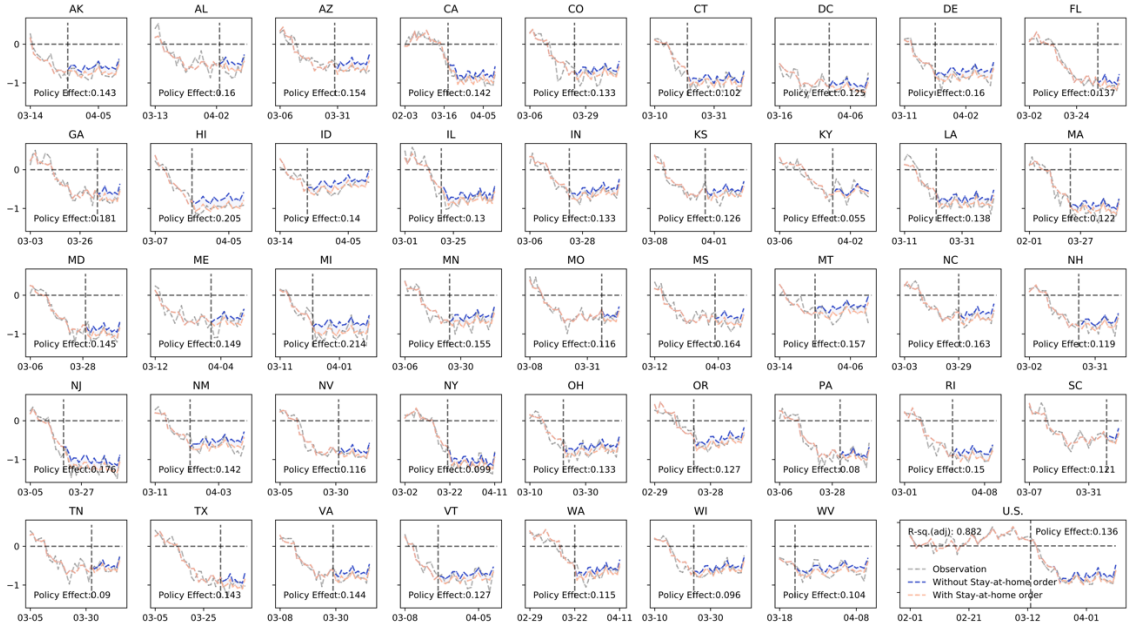

**Figure 3.** Estimated daily trip increase/reduction per person at the National level and for each state and the quantified policy effect of Stay-at-home orders

In each sub-figure, the red curve indicates the model prediction of daily trip increase/reduction per person; the blue curve indicates the model prediction assuming without Stay-at-home orders; Policy Effect indicates the estimated reduction of daily number of trip per person that is solely attributed to the effect of Stay-at-home orders; The vertical dashed line indicates the effective date of the orders.

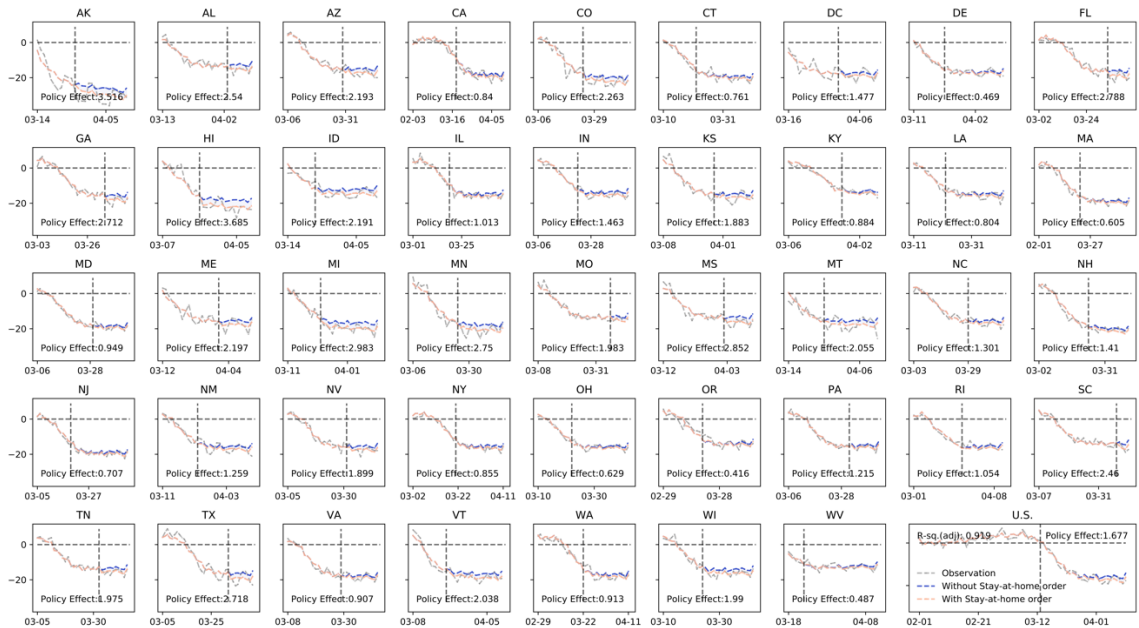

**Figure 4.** Estimated daily person-miles traveled increase/reduction at the National level and for each state and the quantified policy effect of Stay-at-home orders

In each sub-figure, the red curve indicates the model prediction of daily person-miles traveled increase/reduction; the blue curve indicates the model prediction assuming without Stay-at-home orders; Policy Effect indicates the estimated reduction of daily number of trip per person that is solely attributed to the effect of "Stay-at-home" orders; The vertical dashed line indicates the effective date of the orders.

**Acknowledgments**

We would like to thank and acknowledge our partners and data sources in this effort: (1) various mobile device location data providers including a COVID-19 International Data Collaborative led by Cuebiq; (2) Amazon Web Service and its Senior Solutions Architect, Jianjun Xu, for providing cloud computing and technical support; (3) computational algorithms developed and validated in a previous USDOT Federal Highway Administration's Exploratory Advanced Research Program project; and (4) COVID-19 confirmed case data from the Johns Hopkins University Github repository and sociodemographic data from the U.S. Census Bureau.

**Supplementary Information (SI) Appendix**

**Section I. Descriptive Statistics**

Table S1 shows the relative change from the benchmark (i.e., January 2020) in Average Number of Trips per Person and Average Person-Miles Traveled (PMT) per state, three, two and one week before and one week after each "Stay-at-home" order took effect.

- Starting from as early as two weeks before each state officially issued the "Stay-at-home" order, 21 over 43 states experienced more than 10% average PMT drop compared to the January benchmark. And an average of 15% decrease in average PMT (9% decrease in number of trips per person) across all 43 states compared to previous week are observed. These indicate that people were actively taking actions by reducing their travel prior to the issuance of the orders from the state governments.
- Observation on one week before the "Stay-at-home" orders further shows more decrease in PMT, where all 43 states were observed more than 10% PMT drop, and an average of 17% decrease in average PMT (10% decrease in number of trips per person) across all 43 states compared to previous week are observed.
- One week after each state issued its own "Stay-at-home" order, all 43 states experienced more than 30% average PMT drop, and an average of additional 11% decrease in average PMT (6% decrease in number of trips per person) across all 43 states compared to previous week are observed.

Another finding is that the amount decrease in average PMT is larger than that of average number of trips per person, with an average of 6.5%, 13.8% and 18.6% more decrease across all 43 states for two weeks before, one week before and one week after the "Stay-at-Home" orders are issued. One possible reason is that even though people reduce long-distance or commute trips, they tend to do more short-distance activities such as walking their dogs, jogging and other outdoor exercise activities. As most states' "Stay-at-Home" allow outdoor exercises, the results indicate that people do follow the orders to some extent.

**Table S1:** Relative change from the human mobility of January 2020 in Average Number of Trips per Person (# of Trips) and Average Person-Miles Traveled (PMT) per state. Stats for three, two and one week before and one week after each "Stay-at-home" order took effect are shown.

| State | Three weeks before | | Two weeks before | | One week before | | One week after | |
|---|---|---|---|---|---|---|---|---|
| | # of Trips | PMT | # of Trips | PMT | # of Trips | PMT | # of Trips | PMT |
| AL | -1.5% | -7.1% | -12.1% | -24.3% | -12.2% | -27.4% | -16.3% | -32.7% |
| AK | 4.3% | 0.2% | -2.0% | -14.7% | -17.1% | -40.8% | -21.3% | -51.4% |
| AZ | 3.6% | 3.2% | -9.1% | -17.6% | -13.2% | -30.5% | -16.3% | -36.3% |
| CA | 6.4% | 5.1% | 3.8% | 1.9% | -5.6% | -12.0% | -21.3% | -34.0% |
| CO | 7.2% | 3.3% | -1.6% | -8.2% | -17.5% | -33.3% | -21.4% | -44.6% |
| CT | 5.7% | 6.6% | 1.7% | -4.4% | -16.7% | -31.6% | -27.9% | -45.5% |
| DE | 6.2% | 5.8% | 1.2% | -5.4% | -14.8% | -28.9% | -23.2% | -40.4% |
| DC | -8.3% | -8.1% | -25.6% | -38.0% | -32.3% | -55.5% | -33.0% | -58.9% |
| FL | -4.8% | -6.8% | -19.3% | -28.6% | -23.3% | -38.1% | -26.3% | -43.3% |
| GA | -1.7% | -7.1% | -14.8% | -26.0% | -15.2% | -31.0% | -19.4% | -37.8% |
| HI | 4.1% | 0.5% | -0.6% | 1.5% | -14.6% | -27.4% | -26.2% | -51.5% |
| ID | 7.2% | 10.0% | 3.5% | -0.6% | -5.1% | -15.1% | -14.1% | -30.6% |
| IL | 8.7% | 10.3% | 7.7% | 8.1% | -8.3% | -12.1% | -24.4% | -34.8% |
| IN | 7.9% | 9.7% | 1.9% | 1.5% | -11.9% | -20.3% | -17.9% | -36.0% |
| KS | 3.4% | 9.5% | -11.2% | -17.8% | -14.9% | -33.5% | -18.3% | -38.9% |
| KY | 6.7% | 7.5% | -3.7% | -5.1% | -15.4% | -24.9% | -14.2% | -31.2% |
| LA | 6.4% | 7.8% | 4.9% | 2.4% | -11.0% | -19.2% | -19.5% | -31.3% |
| ME | -1.7% | -1.7% | -16.3% | -28.4% | -19.2% | -34.2% | -21.3% | -39.6% |
| MD | 1.5% | -3.5% | -14.8% | -28.7% | -25.8% | -44.1% | -26.6% | -48.5% |
| MA | 5.3% | 5.7% | -1.0% | -6.9% | -21.0% | -35.5% | -29.8% | -49.5% |
| MI | 7.5% | 13.3% | 2.4% | 1.2% | -12.7% | -24.5% | -27.1% | -47.0% |
| MN | 7.7% | 14.5% | -3.1% | -5.2% | -16.3% | -32.8% | -20.0% | -44.0% |
| MS | -0.6% | -0.4% | -13.6% | -23.4% | -12.3% | -25.0% | -17.1% | -31.6% |
| MO | -9.6% | -15.0% | -16.0% | -30.6% | -15.2% | -32.5% | -14.3% | -32.1% |
| MT | 6.0% | 6.4% | -1.5% | -12.7% | -6.8% | -24.6% | -15.1% | -36.9% |
| NV | 0.9% | -4.5% | -18.8% | -26.4% | -21.5% | -34.4% | -22.5% | -39.8% |
| NH | 5.1% | 6.1% | -6.1% | -12.7% | -18.6% | -36.8% | -23.8% | -45.2% |
| NJ | 5.1% | 5.7% | 4.2% | -0.1% | -14.0% | -24.3% | -31.3% | -47.6% |
| NM | 7.4% | 8.0% | 4.7% | 0.5% | -9.1% | -19.5% | -15.7% | -33.1% |
| NY | 5.3% | 5.1% | 2.5% | -0.5% | -15.7% | -23.5% | -32.3% | -45.2% |
| NC | 4.8% | 1.4% | -8.0% | -20.2% | -13.0% | -31.6% | -15.2% | -36.5% |
| OH | 6.7% | 7.3% | 2.7% | 0.0% | -13.5% | -23.3% | -21.3% | -39.4% |
| OR | 6.7% | 7.6% | 3.6% | -2.5% | -4.2% | -14.3% | -17.9% | -33.7% |
| PA | -0.5% | -3.0% | -21.0% | -30.7% | -26.5% | -42.1% | -23.8% | -42.9% |
| RI | 3.9% | 1.7% | -11.2% | -19.3% | -21.0% | -36.7% | -25.9% | -44.2% |
| SC | -9.8% | -20.2% | -11.8% | -27.9% | -10.9% | -32.1% | -11.8% | -34.4% |
| TN | -1.7% | -0.4% | -14.6% | -22.4% | -13.3% | -28.9% | -13.8% | -33.3% |
| TX | -2.6% | 0.2% | -17.0% | -25.4% | -21.4% | -34.1% | -24.1% | -37.7% |
| VT | 3.8% | 14.3% | -1.6% | 4.3% | -16.7% | -30.8% | -25.5% | -43.8% |
| VA | 2.6% | -1.1% | -12.1% | -25.6% | -19.3% | -38.5% | -20.6% | -42.5% |
| WA | 5.9% | 6.9% | 1.2% | -2.2% | -8.7% | -19.5% | -21.6% | -38.3% |
| WV | 6.4% | 6.4% | 2.8% | 0.5% | -11.4% | -20.5% | -16.6% | -31.2% |
| WI | 7.6% | 12.7% | 2.2% | 3.2% | -16.7% | -23.6% | -20.6% | -38.1% |

**Section II. Models**

*2.1. Model Description*

This section provides a detailed description of the GAM we employed to examine the policy effects on human mobility change. GAM is a semi-parametric model with a linear predictor involving a series of additive non-parametric smooth functions of covariates. Compared to the classical ordinary least squares (OLS) regression, GAM is more flexible with fewer assumptions, which is useful when data cannot meet OLS assumptions, such as independence, normality, and homogeneity. Additionally, a noticeable advantage of GAM lies in its capability and flexibility to handle different formats of nonlinear effects (Wood, 2003). By changing the spline functions, various effects can be captured under one model framework, including the random effects, the interaction relationships, and the spatiotemporal autocorrelations.

As a longitudinal analysis with repeated observations over time for each state, the non-independence among the repeated observations and the heterogeneous variability over time should be carefully addressed. Mixed (also named multilevel) models are widely used to handle the panel data (Wolfinger and O'connell, 1993). However, traditional mixed models are linear-based and fail to obtain high performance under data with significant nonlinear fluctuation. Hence, a GAM structure is involved to handle the panel data, with several additive smooth terms besides the linear fixed effect to address the heterogeneous covariance structures. To specific, the additive terms including:

1) Random effects across all states, to capture the unobserved heterogeneity.
2) Interactions between stay-at-home order and state, to capture the varying effect of policies across different states;
3) Time-varying patterns, including an average changing pattern and a seasonally changing pattern (weekly patterns), to fit the autoregressive time series;
4) Spatiotemporal interactions, to capture the spatiotemporal heterogeneity over time across different states.

GAM is estimated using the R '*mgcv*' package (Wood, 2017). Variance components are estimated by the Restricted Maximum Likelihood Estimation (REML), which is widely used in models with random effects. The expression of GAM is shown as follows:

$$T_i^b = \beta_0 + \sum_{k=1}^{K} \beta_{ik} X_{ik} + \sum_{l=1}^{L} f_{il}(X_{il}) + \sum_{r=1}^{R} \sum_{s=1}^{S} \tilde{f}_{ir}(X_{ir}) \times \tilde{f}_{is}(X_{is}) + b \quad \text{S (3.1)}$$

where $T_i^b$ is the vector of the average number of trips per person or average person-miles traveled in state *i* over different days; $\beta_0$ is the overall intercept; $\beta_{ik}$ is the $k^{th}$ coefficient of fixed effects that vary across different states; *K* is the total number of fixed effects; $X_{ik}$ refers to the $k^{th}$ fixed covariate; *L* is the total number of covariates that present nonlinear features; $f_i(.)$ is a low rank isotropic smooth function and $X_{il}$ denotes the $l^{th}$ covariate with nonlinear effects; $X_{ir}$ and $X_{is}$ are the $r^{th}$ pair of interaction covariate; *R* and *S* are the numbers of variables with interactive effects; $\tilde{f}_i(.)$ is an interaction smooth functions with penalties on each null space component; *b* is the random effect vector of a state, and assumed to follow a Gaussian distribution, noted as $N(0, \sigma^2)$; in this study, random effects are parametric terms penalized by a ridge penalty.

*2.2. Variable Description*

Two dependent variables are considered, i.e. the Daily Average Number of Trips Per Person (TPP) and the Daily Average Person-Miles Traveled (PMT), which are used to represent the changes of state-level individual travel frequency. All the dependent variables are the relative value using the corresponding values in January as reference. In other words, they are the increase compared with the same day of the week in January:

$$\check{Y}_d = Y_d - \overline{YR}_{k|W_k=W_d} \qquad \text{S (3.2)}$$

where $\check{Y}_d$ is the relative dependent variables in day $d$; $Y_d$ is the absolute dependent variables in day $d$; $W_d$ is the week of day $d$; $\overline{YR}_{k|W_k=W_d}$ is the average value of TPP or PMT in days belonging to the week $W_d$ in January.

Independent variables include the policy-related features, such as the stay-at-home order with different level of enforcement and the state government approval rate; the cases-related features, such as the daily new cases in the state, the adjacent states, and the nationwide cases; and the temporal variables, such as the time index, the week, and whether is weekend.

The variance inflation factor (VIF) is used to check for multicollinearity, and variables with VIF values greater than 5.0 were excluded. It is worth mentioning highly multicollinearity is observed between the number of new cases and the accumulated cases, and thus the accumulated number of cases is excluded. Similar high multicollinearity is observed between stay-at-home order and COVID-19 Emergency Declaration, and we keep the stay-at-home order in the final models.

The summary of variables is reported in Table S2. The average of TPP and PMT are both negative, indicating the trip frequency and trip miles are both presenting the decreasing trends. The large St.D., on the other hand, implying the changes are heterogeneous across different states.

### 2.3. Modeling Results and Discussions

The results of the two GAMs are shown in Table S3 and S4, respectively. Two components are included: the parametric coefficients, corresponding to the linear fixed effects; and the nonparametric smooth terms, corresponding to the nonlinear effects, random effects (bs='re'), and interaction effects (bs='fs'). Model fit indexes are 0.882 and 0.919 for the two models, indicating that GAMs fit the data well.

*Linear effects* - the stay-at-home orders present significant negative effects on both the number of trips and the person-miles traveled. With the degree of enforcement becoming more severe, the effects of stay-at-home orders on reducing mobility also increase. For case-related variables, the number of nationwide cases is significantly and negatively correlated with both the number of trips and the person-miles traveled. The number of cases in the states, however, only presents a significantly influence on the number of trips, not on the person-miles traveled. For temporal features, the weekend presents significant positive relationships in two models, indicating the reduction of trips on weekends is less than weekdays (i.e. the increment is greater). We also build in a control variable for the effect of governor approval rate and that is deemed insignificant by the model.

*Nonlinear effects* – the estimated degrees of freedom (e.d.f.) are all largely greater than 1.0, suggesting that strong nonlinearities exist. Besides, the interaction terms in all models present P-values smaller than 0.1, implying these nonlinear effects are statistically significant. The fitted results by the spline functions are shown in Figure S1. The values of the vertical axis show the additive effect of the independent variables on the number of trips and the person-miles traveled.

**Table S2: Summary of Variables in the Models**

| Variable | Description | Mean | St.D. | Min. | 50% | Max. |
|---|---|---|---|---|---|---|
| **Dependent Variables** | | | | | | |
| Avg. Number of Trips | Daily Average Number of Trips Per Person | -0.171 | 0.439 | -1.487 | -0.004 | 0.835 |
| Avg. PMT | Daily Average Person-Miles Traveled | -4.367 | 8.900 | -36.003 | -0.526 | 21.072 |
| **Independent Variables** | | | | | | |
| Stay-at-home order | Categorical Variables. 0: No Stay-at-home order (**Reference**); 1: Stay-at-home order issued without penalty or without specifying enforcement; 2: Stay-at-home order issued and enforced with warning, and possible fine for repeated offense; 3: Stay-at-home order issued and enforced with fine and possible jail time | - | - | - | - | - |
| FEMA | *Dummy Variables: 0: No COVID-19 Emergency Declaration; 1: COVID-19 Emergency Declaration issued* | 0.170 | 0.376 | 0.000 | 0.000 | 1.000 |
| New Cases | Daily number of newly confirmed coronavirus cases in the states (1,000) | 0.143 | 0.709 | 0.000 | 0.000 | 11.186 |
| Sum Cases | *Daily number of accumulated confirmed coronavirus cases in the states (1,000)* | 1.386 | 8.428 | 0.000 | 0.002 | 181.029 |
| Adj. New Cases | Daily number of newly confirmed coronavirus cases in adjacent states (1,000) | 0.056 | 0.728 | 0.000 | 0.000 | 13.082 |
| Adj. Sum Cases | *Daily number of accumulated confirmed coronavirus cases in adjacent states (1,000)* | 0.579 | 8.719 | 0.000 | 0.000 | 205.381 |
| National New Cases | Daily number of newly confirmed coronavirus cases in the nation (1,000) | 7.512 | 11.721 | 0.000 | 0.068 | 35.114 |
| Week | The day of week, from 0 (Monday) to 6 (Sunday) | 3.030 | 2.000 | 0.000 | 3.000 | 6.000 |
| Is Weekend | If the day is weekend, 1; else 0 | 0.296 | 0.457 | 0.000 | 0.000 | 1.000 |
| Time Index | The day difference from the current timestamp to 02/01/2020 | 35.271 | 20.564 | 0.000 | 35.000 | 70.000 |
| Approval Rate | State governor approval rate | 3.359 | 12.390 | 0.340 | 0.520 | 59.000 |

*Italic texts: excluded variables due to multicollinearity.*

**Table S3: Estimated GAM Model of Daily Average Number of Trips Per Person**

| Parametric coefficients: | | | | | |
|---|---|---|---|---|---|
| | Estimate | Std. Error | t value | Pr(>\|t\|) | |
| (Intercept) | 0.025 | 0.023 | 1.051 | 0.293 | |
| Stay-at-home order issued without penalty or without specifying enforcement | -0.122 | 0.026 | -4.774 | 0.000 | *** |
| Stay-at-home order issued and enforced with warning, and possible fine for repeated offense | -0.125 | 0.025 | -4.936 | 0.000 | *** |
| Stay-at-home order issued and enforced with fine and possible jail time | -0.167 | 0.024 | -6.865 | 0.000 | *** |
| Daily number of newly confirmed coronavirus cases in the states (1,000) | -0.031 | 0.010 | -3.194 | 0.001 | ** |
| Daily number of newly confirmed coronavirus cases in the adjacent states (1,000) | -0.013 | 0.010 | -1.264 | 0.207 | |
| Daily number of newly confirmed coronavirus cases in the U.S. (1,000) | -0.028 | 0.002 | -12.140 | 0.000 | *** |
| State governor approval rate | -0.001 | 0.001 | -0.402 | 0.688 | |
| Weekend | 0.140 | 0.035 | 3.998 | 0.000 | *** |
| **Approximate significance of smooth terms:** | | | | | |
| | e.d.f | Ref.df | F | P-value | |
| s (Time Index) | 8.839 | 9.000 | 566.409 | 0.000 | *** |
| s (Week) | 4.064 | 5.000 | 61.017 | 0.000 | *** |
| s (State, bs='re') | 0.870 | 48.000 | 0.019 | 0.000 | *** |
| s (Time Index, State, bs='fs') | 108.031 | 498.000 | 4.883 | 0.000 | *** |
| s (Stay at home order, State, bs='re') | 19.430 | 88.000 | 0.325 | 0.000 | *** |
| **Model fit:** | | | | | |
| R-sq.(adj) | | | 0.882 | | |
| Deviance explained | | | 0.887 | | |
| -REML | | | -1418.100 | | |
| Scale est. | | | 0.023 | | |

*Note: '.' p<0.1; '*' p<0.05; '**' p<0.01; '***' p<0.001*

The left figures present the time-dependent random effect (with the dash lines showing the confidence interval), which can be deemed as the impact from other unobserved time-varying factors represented as follows:

- A slight mobility drop is captured by the random effect of both models near 2020/02/15, corresponding to the Presidents' Day weekend.
- A mobility increase is then captured, with a tipping point near 2020/03/07, four days before WHO defined the COVID-19 as a pandemic. In line with others' data findings (e.g. PlaceIQ 2020), we argue this is due to a model-unobserved panic such that people were stocking up goods for the possible lock-down.
- A sharp decrease occurs between 2020/03/07 and 2020/03/22, followed by a dramatic rebound. The rebounding effect for the daily average person-miles traveled is not as steep. This could be due to social distancing fatigue. One explanation is that the increased trips mainly belong to short-distance trips, such as the exercises near home locations. This finding is also in line with other most-recent studies (PlaceIQ, 2020).

The right subplots in Figure S1(a) and S1(b) show the time-varying heterogeneity across different states. With an interaction spline function, these heterogeneities are well captured by the models. Despite all the negative and significant effects from the model variables, states such as DC, NJ, MA, FL, and TX, present additional decreasing trends in heterogeneity, indicating extra caution in these state residents. States such as ID, MT, WY, UT, mostly present increasing trends.

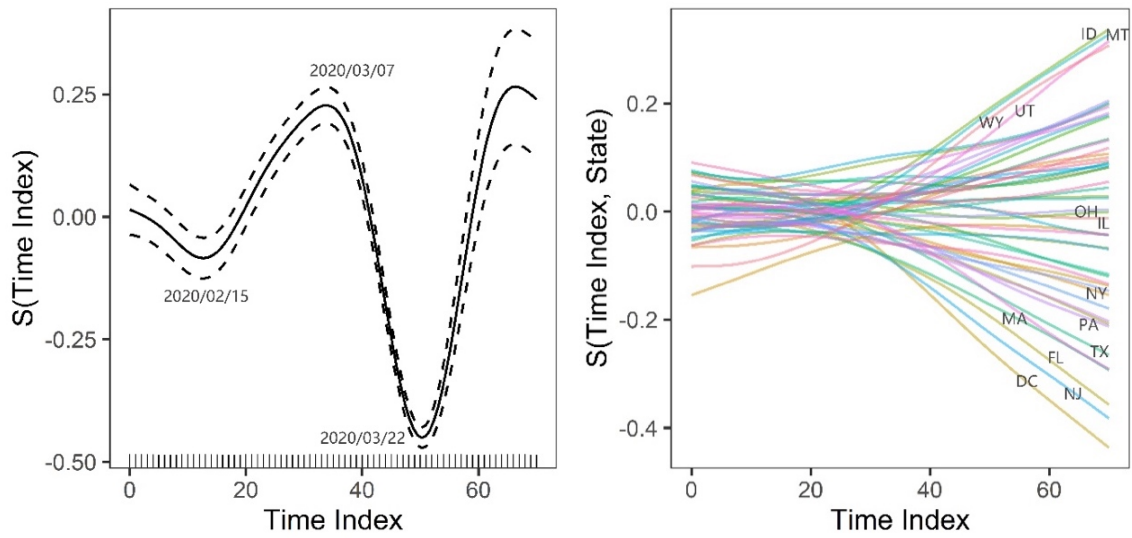

(a) Daily Average Number of Trips Per Person Model (Temporal Effect and State Heterogeneity)

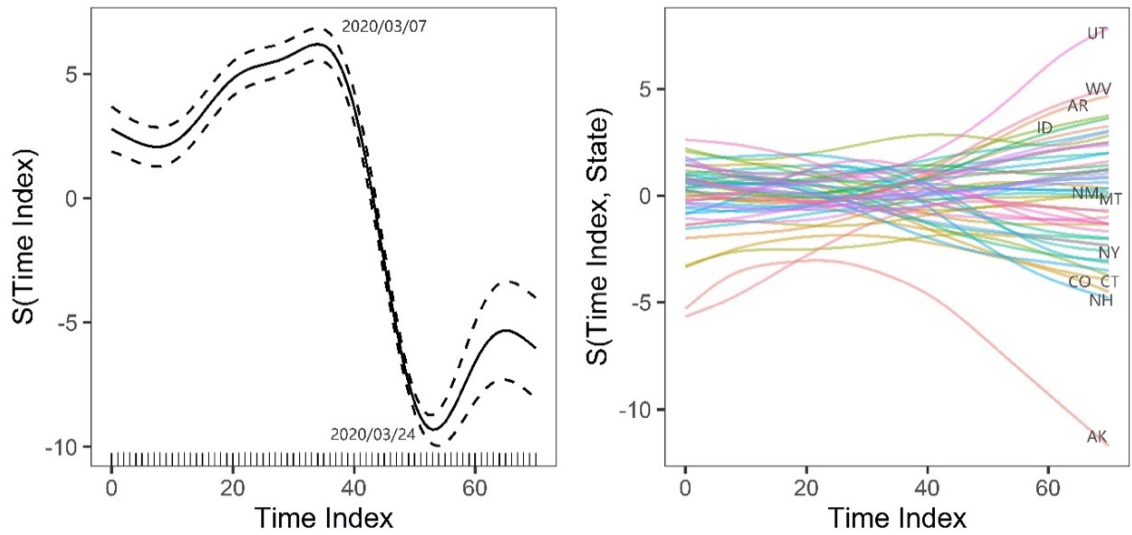

(b) Daily Average Person-Miles Traveled Model (Temporal Effect and State Heterogeneity)

**Figure S1.** Nonlinear temporal effects in Daily Average Number of Trips Per Person Model (a) and Daily Average Person-Miles Traveled Model (b).

**Table S4: Estimated GAM Model of Daily Average Person-Miles Traveled**

| Parametric coefficients: | Estimate | Std. Error | t value | Pr(>|t|) | |
|---|---|---|---|---|---|
| (Intercept) | -2.690 | 0.488 | -5.511 | 0.000 | *** |
| Stay-at-home order issued without penalty or without specifying enforcement | -1.503 | 0.506 | -2.971 | 0.003 | ** |
| Stay-at-home order issued and enforced with warning, and possible fine for repeated offense | -0.883 | 0.503 | -1.754 | 0.080 | . |
| Stay-at-home order issued and enforced with fine and possible jail time | -1.311 | 0.484 | -2.710 | 0.007 | ** |
| Daily number of newly confirmed coronavirus cases in the states (1,000) | 0.304 | 0.187 | 1.629 | 0.457 | |
| Daily number of newly confirmed coronavirus cases in the adjacent states (1,000) | 0.063 | 0.200 | 0.316 | 0.752 | |
| Daily number of newly confirmed coronavirus cases in the U.S. (1,000) | -0.282 | 0.039 | -7.179 | 0.000 | *** |
| State governor approval rate | -0.021 | 0.032 | -0.669 | 0.504 | |
| Weekend | 1.823 | 0.787 | 2.317 | 0.021 | * |
| **Approximate significance of smooth terms:** | e.d.f | Ref.df | F | P-value | |
| s (Time Index) | 8.849 | 9.000 | 490.081 | 0.000 | *** |
| s (Week) | 4.708 | 5.000 | 42.043 | 0.000 | *** |
| s (State, bs='re') | 0.009 | 48.000 | 0.000 | 0.000 | *** |
| s (Time Index, State, bs='fs') | 122.700 | 498.000 | 11.454 | 0.000 | *** |
| s (Stay at home order, State, bs='re') | 27.100 | 88.000 | 0.566 | 0.000 | *** |
| **Model fit:** | | | | | |
| R-sq.(adj) | 0.919 | | | | |
| Deviance explained | 0.923 | | | | |
| -REML | 8634.800 | | | | |
| Scale est. | 6.567 | | | | |

*Note: '.' p<0.1; '*' p<0.05; '**' p<0.01; '***' p<0.001*

### *2.4 Partial Dependence Plot (PDP)*

The various additive nonlinear effects contribute to a high performance of model fit. The coefficients in the linear part only present the average fixed effects, however, the random effects across different states are eliminated. Thus, a partial dependence plot (PDP) method is introduced to examine the effects of stay-at-home orders on travel patterns across different states.

PDP is widely used to interpret black-box models like various machine learning methods (Friedman, 2001). It shows the dependence between the response variable and the predictor, marginalizing over the values of all other predictors. In this study, the partial dependence of stay-at-home order is calculated (see Table S5), serving as the policy effect for each state. The plots of PDP are presented in Figure 3 and 4 and discussed in the main text.

$$P_i = Y_i - \hat{Y}_i \qquad \text{S (3.3)}$$

where $Y_i$ is the predicted number of trips or PMT of the state $i$; $\hat{Y}_i$ is the predicted number of trips or PMT of state $i$ when the value of the stay-at-order variable is set as zero.

**Table S5:** The Model Estimated Policy Effect on Daily Average Trips per Person (TPP) and Daily Person-Miles Traveled (PMT)

| State | Policy effect on TPP | Policy effect on PMT | Rank of TPP | Rank of PMT |
|---|---|---|---|---|
| KY | 0.055 | 0.884 | 1 | 11 |
| PA | 0.080 | 1.215 | 2 | 17 |
| TN | 0.090 | 1.975 | 3 | 25 |
| WI | 0.096 | 1.990 | 4 | 27 |
| NY | 0.099 | 0.855 | 5 | 10 |
| CT | 0.102 | 0.761 | 6 | 7 |
| WV | 0.104 | 0.487 | 7 | 3 |
| WA | 0.115 | 0.913 | 8 | 13 |
| MO | 0.116 | 1.983 | 9 | 26 |
| NV | 0.116 | 1.899 | 10 | 24 |
| NH | 0.119 | 1.410 | 11 | 20 |
| SC | 0.121 | 2.460 | 12 | 34 |
| MA | 0.122 | 0.605 | 13 | 4 |
| DC | 0.125 | 1.477 | 14 | 22 |
| KS | 0.126 | 1.883 | 15 | 23 |
| VT | 0.127 | 2.038 | 16 | 28 |
| OR | 0.127 | 0.416 | 17 | 1 |
| IL | 0.130 | 1.013 | 18 | 15 |
| OH | 0.133 | 0.629 | 19 | 5 |
| IN | 0.133 | 1.463 | 20 | 21 |
| CO | 0.133 | 2.263 | 21 | 33 |
| FL | 0.137 | 2.788 | 22 | 39 |
| LA | 0.138 | 0.804 | 23 | 8 |
| ID | 0.140 | 2.191 | 24 | 30 |
| NM | 0.142 | 1.259 | 25 | 18 |
| CA | 0.142 | 0.841 | 26 | 9 |
| AK | 0.143 | 3.516 | 27 | 42 |
| TX | 0.143 | 2.718 | 28 | 37 |
| VA | 0.144 | 0.907 | 29 | 12 |
| MD | 0.145 | 0.949 | 30 | 14 |
| ME | 0.149 | 2.197 | 31 | 32 |
| RI | 0.150 | 1.054 | 32 | 16 |
| AZ | 0.154 | 2.193 | 33 | 31 |
| MN | 0.155 | 2.750 | 34 | 38 |
| MT | 0.157 | 2.055 | 35 | 29 |
| DE | 0.160 | 0.469 | 36 | 2 |
| AL | 0.160 | 2.540 | 37 | 35 |
| NC | 0.163 | 1.301 | 38 | 19 |
| MS | 0.164 | 2.852 | 39 | 40 |
| NJ | 0.176 | 0.707 | 40 | 6 |
| GA | 0.181 | 2.712 | 41 | 36 |
| HI | 0.205 | 3.685 | 42 | 43 |
| MI | 0.214 | 2.983 | 43 | 41 |